\documentclass{ieeetj}
\usepackage{cite}
\usepackage{gensymb}
\usepackage{siunitx}
\usepackage{amsmath,amssymb,amsfonts}
\usepackage{algorithmic}
\usepackage{graphicx,color}
\usepackage{textcomp}
\usepackage{xcolor}
\usepackage{hyperref}
\hypersetup{hidelinks=true}
\usepackage[ruled,vlined]{algorithm2e}
\def\BibTeX{{\rm B\kern-.05em{\sc i\kern-.025em b}\kern-.08em
    T\kern-.1667em\lower.7ex\hbox{E}\kern-.125emX}}
\AtBeginDocument{\definecolor{tmlcncolor}{cmyk}{0.93,0.59,0.15,0.02}\definecolor{NavyBlue}{RGB}{0,86,125}}
\usepackage[caption=false,font=normalsize,labelfont=sf,textfont=sf]{subfig}

\SetKwInput{KwInput}{Input}                
\SetKwInput{KwOutput}{Output} 
\SetKwInput{Preprocessing}{Preprocessing}
\SetKwInOut{Initialization}{Initialization}
\renewcommand{\Indentp}[1]{%
  \advance\leftskip by #1
  \advance\skiptext by -#1
  \advance\skiprule by #1}%
\renewcommand{\Indp}{\algocf@adjustskipindent\Indentp{\algoskipindent}}
\renewcommand{\Indpp}{\Indentp{1.5em}}%
\renewcommand{\Indm}{\algocf@adjustskipindent\Indentp{-\algoskipindent}}
\renewcommand{\Indmm}{\Indentp{-1.5em}}%
\makeatother

\usepackage{tablefootnote}
\newcommand{\argmin}{\mathop{\mathrm{argmin}}}
\newcommand{\argmax}{\mathop{\mathrm{argmax}}}
\newcommand{\tr}{\mathop{\mathrm{tr}}}
\usepackage{ragged2e}
\usepackage{mathtools}
\usepackage{multirow}
\usepackage{stfloats}
\usepackage{url}
\usepackage{verbatim}
\let\labelindent\relax 
\usepackage{enumitem}
\usepackage{bm}
\usepackage{multirow}
\newtheorem{definition}{Definition}[section]
\usepackage{pifont}
\newcommand{\cmark}{\ding{51}}%

\def\OJlogo{\vspace{-4pt}$<$Society logo(s) and publication title will appear here.$>$}
\def\seclogo{\vspace{10pt}$<$Society logo(s) and publication title will appear here.$>$}

\begin{document}
\receiveddate{XX Month, XXXX}
\reviseddate{XX Month, XXXX}
\accepteddate{XX Month, XXXX}
\publisheddate{XX Month, XXXX}
\currentdate{XX Month, XXXX}

\markboth{}{FUKUHARA {et al.}Time-Varying Graph Signal Estimation among Multiple Sub-Networks}

\title{Time-Varying Graph Signal Estimation among Multiple Sub-Networks}

\author{TSUTAHIRO FUKUHARA\authorrefmark{1}, JUNYA HARA\authorrefmark{1} (Graduate Student Member, IEEE),\\ HIROSHI HIGASHI\authorrefmark{1} (Member, IEEE), AND YUICHI TANAKA\authorrefmark{1} (Senior Member, IEEE)}
\affil{Graduate School of Engineering, Osaka University, Osaka, 565-0871 Japan}
\corresp{Corresponding author: Tsutahiro Fukuhara (email: t.fukuhara@msp-lab.org).}
\authornote{This work is supported in part by JSPS KAKENHI under
Grant 23K26110 and 23K17461, and JST AdCORP under
Grant JPMJKB2307.
}

\begin{abstract}
This paper presents an estimation method for time-varying graph signals among multiple sub-networks.
In many sensor networks, signals observed are associated with nodes (i.e., sensors), and edges of the network represent the inter-node connectivity. 
For a large sensor network, measuring signal values at all nodes over time requires huge resources, particularly in terms of energy consumption.
To alleviate the issue, we consider a scenario that a sub-network, i.e., cluster, from the whole network is extracted and an intra-cluster analysis is performed based on the statistics in the cluster. The statistics are then utilized to estimate signal values in another cluster. This leads to the requirement for transferring a set of parameters of the sub-network to the others, while the numbers of nodes in the clusters are typically different. In this paper, we propose a cooperative Kalman filter between two sub-networks. The proposed method alternately estimates signals in time between two sub-networks. We formulate a state-space model in the source cluster and transfer it to the target cluster on the basis of optimal transport. 
In the signal estimation experiments of synthetic and real-world signals, we validate the effectiveness of the proposed method.
\end{abstract}



\maketitle

\section{INTRODUCTION}
Sensor networks are ubiquitous \cite{chong2003sensor} where their nodes observe the signals of the corresponding sensors and their edges represent the inter-node connectivity.
Their applications are broad, including environmental data collection, bottleneck detection in traffic networks, and leakage detection in infrastructure networks \cite{lee2011discovering,martini2015automatic,gabrys2000general}.

A network often has sub-networks, each having similar statistics called a \textit{community} or \textit{cluster}. 
In many clustered sensor networks, sensor data in one cluster is affected by those obtained at the previous time instances, and they could impact on those in different clusters.
For example, temperature and/or air pressure in one area often affect to those in another one.
Under the circumstances, to correctly capture the evolution of sensor data, one needs to predict the current sensor data in one cluster reflecting the impacts from the sensor data in different clusters as well as those at the previous time instances.
Therefore, predictive control (PC) for networks among clusters is crucial and has been extensively studied in many application fields \cite{van_der_merwe_square-root_2001,clarke1989properties,kouvaritakis2016model,maciejowski2007predictive,sutton1981toward}.

Kalman filter is the most popular PC method for networks \cite{sagi2023extended,ramezani2018joint,bishop2001introduction,li1989state}. 
It linearly tracks and estimates \textit{time-varying} signals on a \textit{static} network by minimizing the mean squared error (MSE).
The system of Kalman filter is modeled by a state-space model where the Kalman estimator is performed in two steps: 
\begin{enumerate}
    \item  \textit{Prediction step} calculates the prior estimation of the signals based on their statistics and then derives the Kalman gain \textit{from} the prior estimation.
    \item  \textit{Update step} obtains the posterior estimation of the signals by applying the Kalman gain \textit{to} the prior estimation. 
\end{enumerate}

We often encounter large networks,
however, observing all of their signal values at every time instance may be costly in terms of storage burdens and energy consumption, which may also shorten the lifetime of sensors \cite{zhu2015green}.
For reducing the sensing cost, we consider the following scenario: One cluster from the network is first extracted, and then an intra-cluster analysis based on the statistics in the cluster is performed \cite{natali_learning_2022,pan_survey_2010}. 
If the two clusters are similar\footnote{The similarity of the two clusters is later discussed in Section \ref{sec:CKF}--\ref{sec:signal-model}.}, we can estimate signal values in another cluster if the obtained statistics can be transferred to the other one. 
While this could lead to the reduction of the sensing cost,
we need to develop a method for transferring statistics, i.e., a set of parameters in Kalman filter, between clusters where they typically have different cluster sizes.

\begin{figure}
    \centering
    \includegraphics[width = 0.9 \linewidth]{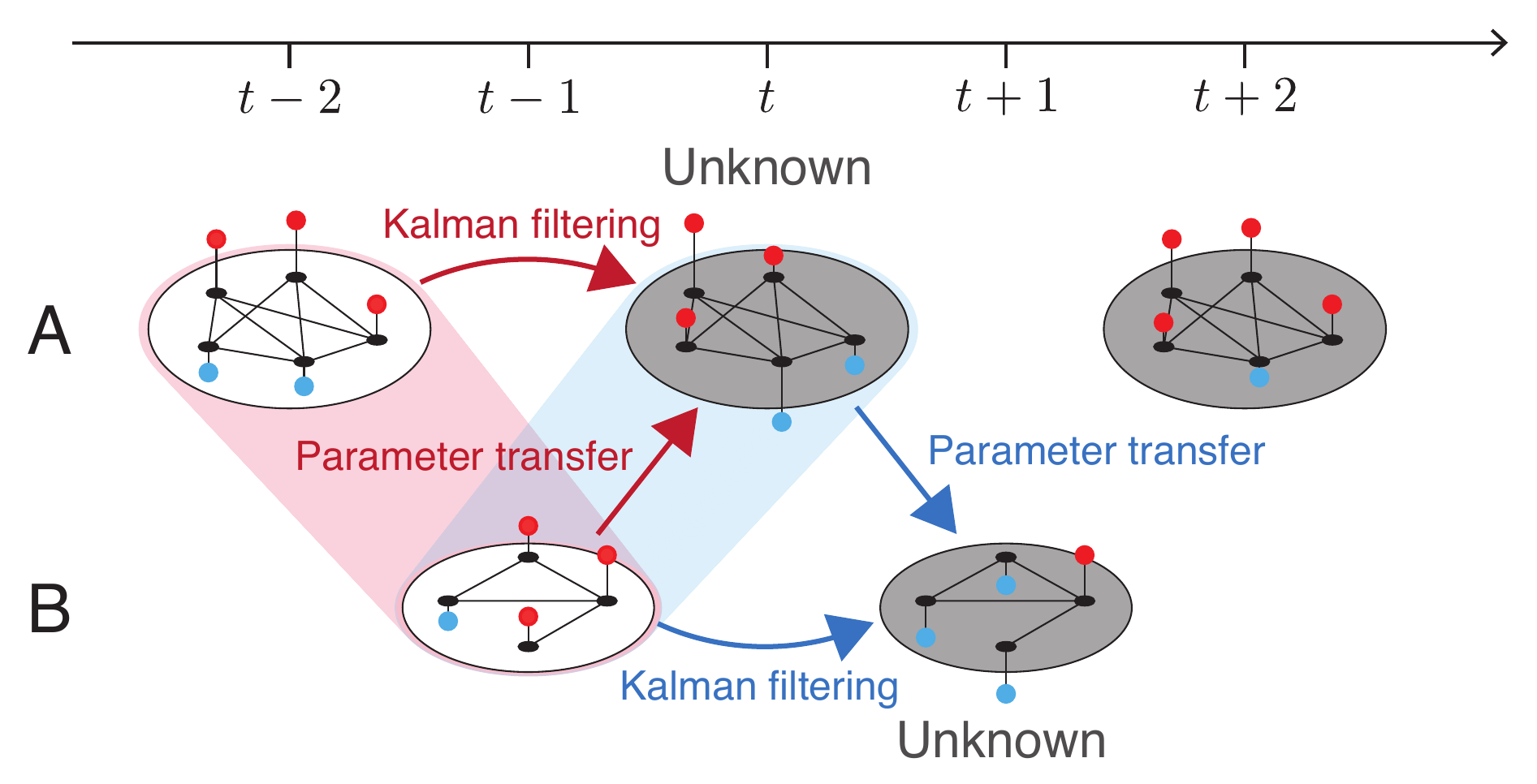}
    \caption{Overview of a cooperative Kalman filter for time-varying graph signals on two static subgraphs. Colored areas denote the set of the source and target at each time instance.}
    \label{fig:overview}
\end{figure}

\IEEEpubidadjcol

In this paper, we propose a cooperative Kalman filter for estimating time-varying signals among multiple sub-networks, i.e., subgraphs. In the estimation, \textit{graph signals}, defined as signals whose domain is a set of nodes in a graph \cite{shuman2013emerging,ortega_graph_2018}, are used to model the signals observed in a network. We illustrate the overview of the proposed method in Fig.~\ref{fig:overview}: The proposed Kalman filter performs its estimation alternately in time between two subgraphs\footnote{While this paper focuses on estimation between two subgraphs for simplicity, it could be generalized to three or more subgraphs.}.
That is, the two subgraphs and signals on them work as the source and target alternately.
The set of parameters in the source is transferred to the target and then it is used in the Kalman estimator, which estimates the target graph signals.

In this paper, first, we propose a \textit{cyclic} graph wide sense stationarity (CGWSS) of time-varying graph signals.
CGWSS is an extension of the widely-studied GWSS \cite{perraudin_stationary_2017,hara_graph_2023,marques_stationary_2017} to the case where its power spectral density (PSD) changes periodically over time. 
Second, we formulate the \textit{state-space model} in the source domain. 
The state equation is derived based on optimal transport, which determines an efficient mapping between two sets of random signals \cite{villani2009optimal,delon_wasserstein-type_2020}.
In our model, we utilize optimal transport as the signal transition filter from the previous signal to the current signal in a subgraph.
Finally, we transfer the state-space model of the source into the target subgraph by reflecting their statistics.
In this transfer, the statistics between the two subgraphs will be different since the numbers of nodes in the subgraphs are different. Especially, under the CGWSS assumption, the two subgraphs will differ in their PSDs. 
To compensate for the gap of the PSDs, we utilize a transfer method based on the Bayesian inference \cite{yamada_graph_2022}.
As a result, we obtain the state-space model of the target and derive the corresponding control laws.
Since the proposed method estimates the graph signals between two subgraphs alternately and asynchronously, we can reduce the sampling frequency for each subgraph in half.

We summarize the differences between the proposed method and existing signal estimation methods in Table \ref{tab:Difference}.
As shown in Table~\ref{tab:Difference}, standard Kalman filter \cite{bishop2001introduction} performs estimation only in one graph and its filter cannot be transferred to the other graph.
While the proposed method and transferred graph Wiener filter (TGWF) \cite{yamada_graph_2022} estimate graph signals between two graphs,
there are two major differences as follows:
\begin{enumerate}
    \item 
    TGWF stimates the current signal by only using the \textit{space} equation in the state-space model.
    In contrast, the proposed method utilizes the \textit{state-space} model.
    \item 
    Graph signals considered in \cite{yamada_graph_2022} conform to \textit{static} GWSS \cite{perraudin_stationary_2017}. In contrast, the proposed method assumes CGWSS as a \textit{time-varying} version of GWSS.
\end{enumerate}
As can be seen from the differences above, the proposed method naturally extends the existing approaches to the estimation of time-varying graph signals among multiple subgraphs.

Our experiments on synthetic and real data demonstrate that the proposed method effectively estimates time-varying graph signals between two static subgraphs.

\begin{table}[t]
    \centering 
    \renewcommand{\arraystretch}{0.82}
    \caption{Main differences between existing methods and proposed method. TGWF and SKF denote the transferred graph Wiener filter \cite{yamada_graph_2022} and standard Kalman filter \cite{bishop2001introduction}.}
    \begin{tabular}{c|c|c|c}
    \hline
         & Proposed filter & TGWF in \cite{yamada_graph_2022} & SKF in \cite{bishop2001introduction} \\\hline\hline
   Target graph & Two  & Two & One \\ \hline
   Signal evolution & Cyclic  & Static & Time-Varying\\ \hline
   Stationarity & CGWSS  & GWSS &  Non-stationary \\ \hline
   System model & State-space  & Space & State-space \\\hline
    \end{tabular}
    \label{tab:Difference}
\end{table}

The remainder of this paper is organized as follows.
Section \ref{sec:pre} introduces the basics of graph signals and graph wide-sense stationarity for random graph signals. Then, we also review Kalman filter and optimal transport, which are building blocks of our framework.
Section~\ref{sec:related} shows an existing graph filter transfer method, which is mainly related to our method.
In Section~\ref{sec:CKF}, we derive a cooperative Kalman filter for estimating time-varying graph signals among multiple subgraphs.
Signal estimation experiments for synthetic and real-world data are demonstrated in Section \ref{sec:exp}. Section \ref{sec:conclude} concludes this paper. 

\textit{Notation}: We denote the $i$th element of a vector $x$ by $[\mathbf{x}]_i$ and denote the $(n,m)$-element of a matrix $\mathbf{X}$ by $[\mathbf{X}]_{nm}$. 

\section{PRELIMINARIES}\label{sec:pre}

In this section, we introduce preliminaries of random signals defined on a graph, i.e., graph signals and review several existing studies on signal estimation. First, we review the basics of graph signal processing (GSP). Second, we define the notion of a graph wide sense stationarity (GWSS). Then, we introduce the standard (linear) Kalman filter and optimal transport, which are related to the main part of our framework.

\subsection{Basics of Graph Signal Processing}
A weighted undirected graph is denoted by $\mathcal{G}=(\mathcal{V},\mathcal{E})$, in which $\mathcal{V}$ and $\mathcal{E}$ are sets of nodes and edges, respectively.
The number of nodes and edges are denoted by $N=|\mathcal{V}|$ and $E=|\mathcal{E}|$, respectively. 
We use a weighted adjacency matrix $\mathbf{W}$ for representing the connection between nodes, where its $(m,n)$-element $[\mathbf{W}]_{mn}\geq 0$ is the edge weight between the $m$th and $n$th nodes; $[\mathbf{W}]_{mn}=0$ for unconnected nodes. 
The degree matrix $\mathbf{D}$ is a diagonal matrix whose each element is defined as $[\mathbf{D}]_{mm}=\sum_n [\mathbf{W}]_{mn}$. Using $\mathbf{D}$ and $\mathbf{W}$, the graph Laplacian is given by $\mathbf{L}=\mathbf{D}-\mathbf{W}$. 
A graph signal $\mathbf{x}\in\mathbb{R}^{N}$ is defined as $\mathbf{x}:\mathcal{V}\to\mathbb{R}^N$ where $[\mathbf{x}]_n$ corresponds to the signal value at the $n$th node.
Since $\mathbf{L}$ is a real symmetric matrix, it has orthogonal eigenvectors and can be diagonalized as $\mathbf{L}=\mathbf{U}\mathbf{\Lambda}\mathbf{U}^\top$,
where $\mathbf{U}=[\mathbf{u}_{0},\mathbf{u}_{1},\dots,\mathbf{u}_{N-1}]$ is a matrix whose $i$th column is the eigenvector $\mathbf{u}_{i}$ and $\mathbf{\Lambda}=\textrm{diag}(\lambda_{0},\lambda_{1},\ldots,\lambda_{N-1})$ is their diagonal eigenvalue matrix.
Without loss of generality, we can assume $0 = \lambda_{0}\le\lambda_{1},\dots,\lambda_{N-1}=\lambda_{\max}$ since $\mathbf{L}$ is a positive semidefinite matrix \cite{shuman2013emerging}.
In GSP, $\lambda_i$ is referred to as a \textit{graph frequency}. Then, spectra of $\mathbf{x}$ in the graph frequency domain are defined as $\hat{\mathbf{x}}=\mathbf{U}^\top\mathbf{x}$: It is called \textit{graph Fourier transform} \cite{ortega_graph_2018}.

\subsection{Graph Wide Sense Stationarity}
Here, we describe the stationarity of a random signal $\mathbf{x}\in\mathbb{R}^N$ on the graph $\mathcal{G}$.
The graph signal $\mathbf{x}$ is a graph wide sense stationary (GWSS) process if the following two conditions are satisfied:

\begin{definition}[GWSS \cite{perraudin_stationary_2017}]\label{def:gwss}Let $\mathbf{x}$ be a random signal on graph $\mathcal{G}$. The signal $\mathbf{x}$ follows a GWSS process if and only if the following conditions are satisfied:
\begin{subequations}
\begin{align}
    \mathbb{E}\left[\mathbf{x}\right]& =\mu=\mathrm{const},\label{eq:gwss1}\\
    \mathbb{E}\left[\mathbf{x}\mathbf{x}^\top\right]& =\mathbf{\Sigma}=\mathbf{U}\mathrm{diag}(\mathbf{p})\mathbf{U}^\top,\label{eq:gwss2}
\end{align}
\end{subequations}
where $\mathbf{\Sigma}$ and  $\mathbf{p}$ are referred to as the covariance matrix and power spectral density (PSD), respectively, and $\mu$ indicates the mean. 
\end{definition}

In this paper, we assume GWSS signals $\mathbf{x}$ are ergodic: The ensemble mean in Definition~\ref{def:gwss} is identical to the temporal mean. Later, we consider a time-varying version of the GWSS process, whose PSD varies periodically over time. 

\subsection{Kalman Filter}\label{sec:kalman-filter}

Kalman filter is an online algorithm for sequentially tracking and estimating time-varying signals from given observations \cite{bishop2001introduction,anderson2005optimal,mesbah2016stochastic,sagi2023extended,ramezani2018joint}.
In the following, we revisit the system of the standard Kalman filter and its control laws.

\subsubsection{System of Kalman Filter}
Let 
the subscript $t$ indicates a time instance. 
The system of Kalman filter is modeled by a state-space model. Generally, it is formulated as follows \cite{bishop2001introduction}:
\begin{subequations}
\begin{align}
    \mathbf{x}_{t}=&\mathbf{A}\mathbf{x}_{t-1}+\mathbf{B}\mathbf{u}_{t-1}+\mathbf{v}_{t},\quad\mathbf{v}_{t}\sim\mathcal{N}(0,\sigma_{v}^2\mathbf{I}), \label{eq:kalman_model-state1}\\
    \mathbf{y}_{t}=&\mathbf{C}\mathbf{x}_{t}+\mathbf{w}_{t},\quad \mathbf{w}_{t}\sim\mathcal{N}(0,\sigma_{w}^2\mathbf{I}),\label{eq:kalman_model-observe1}
\end{align}
\end{subequations}
where $\mathbf{x}_t,\ \mathbf{y}_t\in\mathbb{R}^{M}$ represent the current signal and its observation, respectively, and $\mathbf{u}\in\mathbb{R}^{N}$ indicates the control input.
In \eqref{eq:kalman_model-state1} and \eqref{eq:kalman_model-observe1}, the matrices $\mathbf{A}\in\mathbb{R}^{N\times N}$, 
$\mathbf{B}\in\mathbb{R}^{N\times N}$, and $\mathbf{C}\in\mathbb{R}^{M\times N}$ represent the signal transition, input, and observation matrices, respectively.
The vector $\mathbf{v}_t\in\mathbb{R}^{N}$ represents system noise, and
$\mathbf{w}_t\in\mathbb{R}^{M}$ denotes observation noise, conforming to white Gaussian noise with the standard deviations $\sigma_{v}$ and $\sigma_{w}$, respectively. 
Note that \eqref{eq:kalman_model-state1} describes the transition of the signals over time and \eqref{eq:kalman_model-observe1} represents the observation model at $t$.

In Kalman filtering, the objective is to minimize the mean squared error (MSE) of the estimated signal for all $t$. The optimization problem is thus formulated as:
    \begin{equation}
        \min_{\tilde{\mathbf{x}}_{t}} \mathbb{E}[\|\mathbf{x}_t-\tilde{\mathbf{x}}_{t}\|_2^2],\label{eq:kalman_prob}
    \end{equation}
where $\tilde{\mathbf{x}}_{t}$ represents the estimated signal at $t$. 

We next show the control laws of Kalman filter to estimate the current signal $\tilde{\mathbf{x}}_{t}$ in \eqref{eq:kalman_prob}.

\subsubsection{Control Laws of Kalman Filter}
Hereafter, we denote the estimated signal conditioned on observations up to $t$, i.e., the prior estimation, by $\tilde{\bm{x}}_{t|t-1}$ and denote the posterior one by $\tilde{\bm{x}}_t$.
Kalman filter is performed sequentially with two control laws: 1) prediction and 2) update.
For simplicity, we omit their detailed derivation but they are obtained by solving \eqref{eq:kalman_prob}.
The control laws are given as follows \cite{bishop2001introduction}:
\begin{description}\label{eq:CKF}
    \item[Prediction step]\text{}
        \begin{enumerate}
            \item Calculating the prior estimation.
            \begin{equation}
                \tilde{\mathbf{x}}_{t|t-1}=\mathbf{A}\tilde{\mathbf{x}}_{t-1}+\mathbf{B}\mathbf{u}_{t-1},\label{eq:pre_state}
            \end{equation}
            \item Determining the error covariance matrix of the prior signal estimation.
            \begin{equation}
                \mathbf{P}_{t|t-1}=\mathbf{A}\mathbf{P}_{t-1}\mathbf{A}^\top+\sigma_v^2\mathbf{I},\label{eq:pre_cov}
            \end{equation}
            where $\mathbf{P}_{t-1}$ is the posterior error covariance at $t-1$.
            \item Deriving the optimal filter (Kalman gain) in  \eqref{eq:kalman_prob} from observations up to $t-1$.
            \begin{equation}
                \mathbf{K}_{t|t-1}=\mathbf{P}_{t|t-1}\mathbf{C}^\top(\mathbf{C}\mathbf{P}_{t|t-1}\mathbf{C}^\top+\sigma_w^2\mathbf{I})^{-1},\label{eq:kalman_gain}
            \end{equation}
        \end{enumerate}
    \item[Update step]\text{}
        \begin{enumerate}
            \item Estimating the current signal using the Kalman gain.
            \begin{equation}
                \tilde{\mathbf{x}}_{t}=\tilde{\mathbf{x}}_{t|t-1}+\mathbf{K}_{t|t-1}(\mathbf{y}_{t}-\mathbf{C}\tilde{\mathbf{x}}_{t|t-1}),\label{eq:kalman_est}
            \end{equation}
            \item Updating the posterior error covariance matrix.
            \begin{equation}
                \mathbf{P}_{t}=(\mathbf{I}-\mathbf{K}_{t|t-1}\mathbf{C})\mathbf{P}_{t|t-1}.\label{eq:post_cov}
            \end{equation}
            \item Returning to the prediction step with $t\leftarrow t+1$.
        \end{enumerate}
\end{description}
\noindent At the initial time instance, $t=1$,  $\mathbf{P}_{t-1}$ in \eqref{eq:pre_cov} is usually set to the scaled identity matrix, $\mathbf{P}_{0}=\delta\mathbf{I}$, where the $\delta>0$ is a scaling factor.


\subsection{Optimal Transport}\label{sec:OT}

Optimal transport is a mathematical tool to determine the most efficient mapping between two sets of random signals having different probability distributions \cite{villani2009optimal,santambrogio2015optimal}.

Let us consider the transport from an input signal $\mathbf{x}_{1}\sim\alpha$ to a subsequent signal $\mathbf{x}_{2}\sim\beta$, where $\alpha$ and $\beta$ are two different probabilistic measures on $\mathbb{R}^N$.
Then, the optimal transport seeks the assignment $\gamma$ such that it minimizes some transport cost from $\mathbf{x}_1$ to $\mathbf{x}_2$.
Given a transport cost function $c:\mathbb{R}^N\times\mathbb{R}^N\to\mathbb{R}_+$, the optimal transport problem can be formulated as:
\begin{equation}
    \inf_{\gamma\in\Pi(\alpha,\beta)} \mathop{\mathbb{E}}_{\mathbf{x}_1\sim\alpha,T(\mathbf{x}_1)\sim\beta}c(\mathbf{x}_1,T(\mathbf{x}_1))\quad\text{s.t. }
    T_\#\alpha=\beta,\label{eq:ot_primal}
\end{equation}
where $\mathbf{x}_2=T(\mathbf{x}_1)$ and $\Pi$ represents a subset of joint distributions on $\mathbb{R}^N\times\mathbb{R}^N$.
The mapping $T:\mathbb{R}^N\to\mathbb{R}^N$ is referred to as the optimal transport map, and $T_\#\alpha$ denotes the push-forward measure of $\alpha$ under $T$ \cite{santambrogio2015optimal}.

The cost function often employs the $\ell_p$-norm.
If the $\ell_p$-norm is selected as the cost, the minimum value in \eqref{eq:ot_primal} is referred to as the 
$p$-Wasserstein distance \cite{villani2009wasserstein,panaretos2019statistical}.
 Obtaining the optimal transport map $T$ analytically from \eqref{eq:ot_primal} is generally challenging due to its non-uniqueness \cite{delon_wasserstein-type_2020}.
 Nevertheless, if both of $\alpha$ and $\beta$ are Gaussian distributions, i.e., $\alpha=\mathcal{N}(\bm{\mu}_{1},\mathbf{\Sigma}_{1})$ and $\beta=\mathcal{N}(\bm{\mu}_{2},\mathbf{\Sigma}_{2})$, the unique solution is given by 2-Wasserstein distance $W_2^2(\alpha,\beta)$\cite{panaretos2019statistical}:
\begin{equation}\label{eq:Wasserstein}
\begin{split}
    &W_2^2(\alpha,\beta)=\\
    &{\|\bm{\mu}_{1}-\bm{\mu}_{2}\|}_2^2    +\tr\left(\mathbf{\Sigma}_{1}+\mathbf{\Sigma}_{2}-2\left(\mathbf{\Sigma}_{1}^{1/2}\mathbf{\Sigma}_2
    \mathbf{\Sigma}_{1}^{1/2}\right)^{1/2}\right),
\end{split}
\end{equation}
where $^{1/2}$ is the square root of a matrix \cite{bjorck1983schur}.
Simultaneously, the optimal transport $T$ is obtained \cite{delon_wasserstein-type_2020} as 
\begin{equation}
    T(\mathbf{x}_1)=\bm{\mu}_{2}+\mathbf{\Sigma}_{1}^{-1/2}(\mathbf{\Sigma}_{1}^{1/2}\mathbf{\Sigma}_{2}\mathbf{\Sigma}_{1}^{1/2})^{1/2}\mathbf{\Sigma}_{1}^{-1/2}(\mathbf{x}_{1}-\bm{\mu}_{1}),\\\label{eq:special_ot}
\end{equation}
where we suppose that $\mathbf{\Sigma}_{1}$ is invertible.



\section{Graph Filter Transfer}\label{sec:related}
In this section, we review an existing study on graph filter transfer \cite{yamada_graph_2022}, which is mainly related to the proposed method.
The graph filter transfer aims to transfer parameters of a graph filter for the source graph to the target one having different (but similar) statistics. Specifically, we introduce the method employed for probabilistic filters, such as the graph Wiener filter, Kalman filter, and Bayesian filter \cite{yamada_graph_2022}.

Hereafter, we denote the source and target domains by src and trg, respectively. Let the subscript $\text{dom}\in\{\text{src},\text{trg}\}$ represent one of the two graphs, and $\mathbf{x}_{\text{dom},t}$ and $\mathbf{y}_{\text{dom},t}$ denote an unknown graph signal at time $t$ and its corresponding observation defined in \eqref{eq:kalman_model-observe1}. 
Now, we consider estimating $\mathbf{x}_{\text{src},t}$ using a set of $K$ historical observations, denoted by $\mathbf{Y}_{\text{src},t}=[\mathbf{y}_{\text{src},t},\ldots,\mathbf{y}_{\text{src},t-K+1}]$. We assume that the original graph signal $\mathbf{x}_{\text{dom},t}$ satisfies the GWSS conditions in Definition \ref{def:gwss}.
Under the assumption, the probabilistic filters are generally formulated through maximum a posteriori (MAP) estimation \cite{perraudin_stationary_2017,yamada_graph_2022}, which is expressed as: 
\begin{equation}
\tilde{\mathbf{x}}_{\text{src},t}=\argmax_{\mathbf{x}_{\text{src},t}} \mathcal{P}(\mathbf{x}_{\text{src},t}|\mathbf{Y}_{\text{src},t},\mathbf{p}_{\text{src}}),\label{eq:map_est}
\end{equation} 
where $\mathcal{P}$ denotes a probability distribution function, and $\mathbf{p}_{\text{src}}$ represents the PSD of $\mathbf{x}_{\text{src},t}$. Since $\mathbf{p}_{\text{src}}$ governs the behavior of the estimator in \eqref{eq:map_est}, transferring \eqref{eq:map_est} to the target domain comes down to adjusting $\mathbf{p}_{\text{src}}$ to adapt the target graph. This corresponds to estimating  $\mathbf{p}_{\text{trg}}$ from $\mathbf{p}_{\text{src}}$.

\begin{figure}
    \centering
    \includegraphics[width=0.76\linewidth]{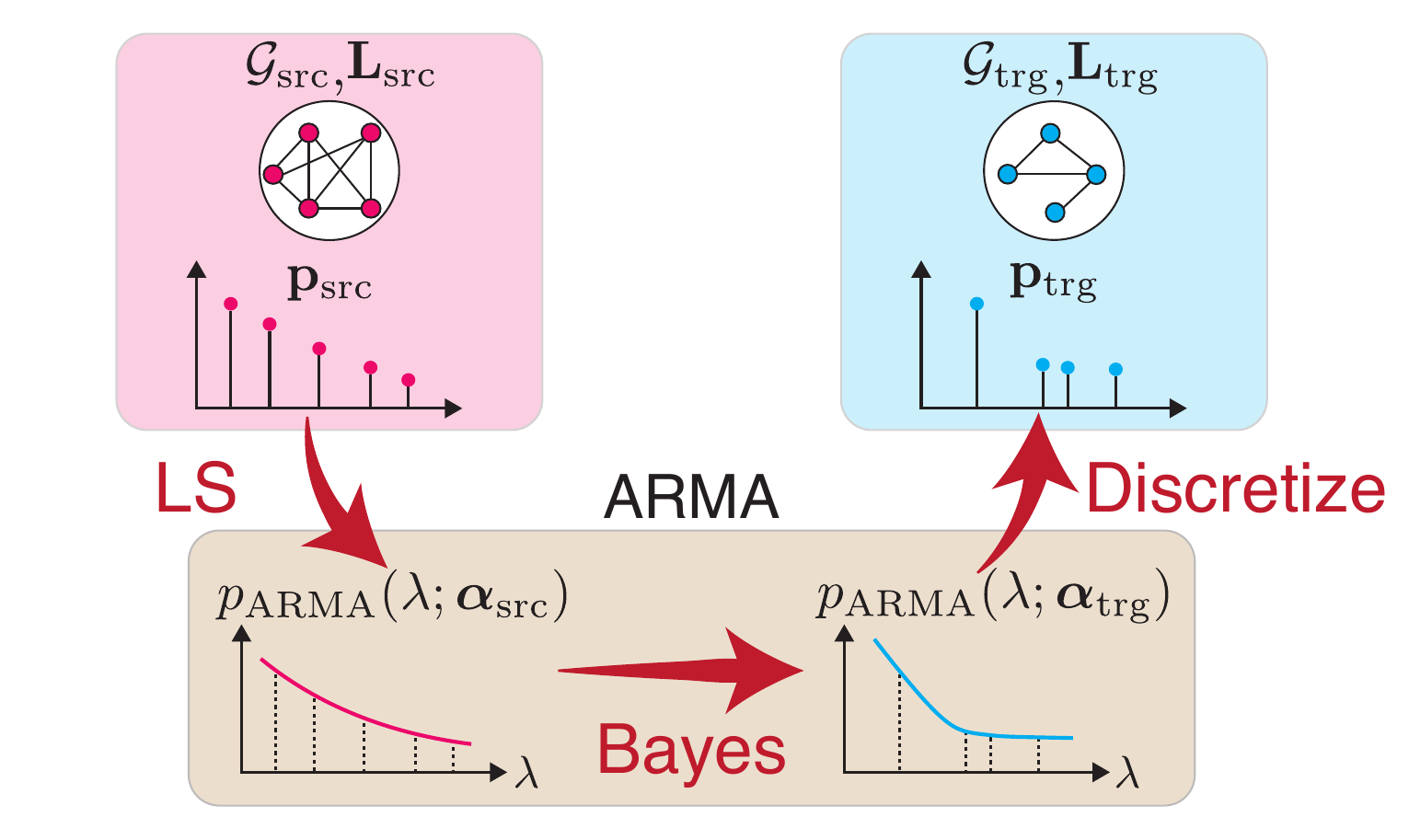}
    \caption{Overview of a graph filter transfer method for the PSD estimation in the target graph in \cite{yamada_graph_2022}.}
    \label{fig:filter_transfer}
\end{figure}

In general, the eigenvalue distributions of different graph variation operators are different.
This discrepancy presents a challenge of estimating $\mathbf{p}_{\text{trg}}$ directly from $\mathbf{p}_{\text{src}}$.
To address the issue, a graph filter transfer method in \cite{yamada_graph_2022} is performed in the following three steps, as visualized in Fig.~\ref{fig:filter_transfer}:

\begin{enumerate}
    \item  Approximating $\mathbf{p}_{\text{src}}$ as a \textit{continuous} ARMA graph filter \cite{isufi_autoregressive_2017} based on the least-squares (LS).
    The approximated PSD is denoted by $p_{\text{ARMA}}(\lambda;\bm{\alpha}_\text{src})$, where $\lambda\in[0,+\infty)$ and $\bm{\alpha}_{\text{dom}}$ represents a sequence of parameters of the ARMA graph filter.
    \item  Adapting $\bm{\alpha}_\text{src}$ in $p_{\text{ARMA}}(\lambda;\bm{\alpha}_\text{src})$ to the target graph using Bayesian inference \cite{kroizer_bayesian_2022}.
    The adapted parameters in the target graph is represented as $\bm{\alpha}_\text{trg}$.
    \item  Discretizing $p_{\text{ARMA}}(\lambda;\bm{\alpha}_\text{trg})$ according to the eigenvalues of $\mathbf{L}_{\text{trg}}$ and estimating $\mathbf{p}_{\text{trg}}$ \cite{kay1993statistical}.
\end{enumerate}
In the three steps, $\mathbf{p}_{\text{trg}}$ is indirectly estimated from $\mathbf{p}_{\text{src}}$ through the parameters of the ARMA filter. 
This approach can appropriately transfer a probabilistic filter from the source graph to the target graph having different eigenvalue distributions.
Please see \cite{yamada_graph_2022} for more details.

While the existing methods only consider signals on the single cluster \cite{bishop2001introduction} or static signals \cite{yamada_graph_2022}, sensor data are typically time-varying and have inter-cluster relationships.
In the following section, we introduce a signal estimation method suitable for the situation.



\section{COOPERATIVE KALMAN FILTER}\label{sec:CKF}

In this section, we introduce our proposed method, a graph filter transfer method for time-varying graph signals using Kalman filter.
Initially, we design a stochastic signal model for time-varying graph signals and then formulate a state-space model. We derive Kalman filter from the state-space model and the graph filter transfer for the target subgraph. 

\subsection{Signal Model}\label{sec:signal-model}


Here, we assume that $\mathbf{x}_{\text{dom}}$ satisfies the conditions defined below:

\begin{definition}[Cyclic Graph Wide Sense Stationary]\label{def:cgwss}Let $\mathbf{x}_t$ be a random signal on $\mathcal{G}$ at the time instance $t$. 
The signal $\mathbf{x}_t$ is a cyclic graph wide sense stationary (CGWSS) process if and only if the following conditions are satisfied:
\begin{subequations}
\begin{align}
    \mathbb{E}\!\left[\mathbf{x}_{t\,(\mathrm{mod}\,P)}\right]\!& =\mu_{t\,(\mathrm{mod}\,P)}\!
     =\mathrm{const},
    \label{eq:cgwss3}\\
    \mathbb{E}\!\left[\mathbf{x}_{t\,(\mathrm{mod}\,P)}\mathbf{x}_{t\,(\mathrm{mod}\,P)}^\top\right]\!& =\mathbf{\Sigma}_{t\,(\mathrm{mod }P)}\!=\mathbf{U}\mathrm{diag}(\mathbf{p}_{t\,(\mathrm{mod}\,P)})\mathbf{U}^\top,
    \label{eq:cgwss4}
\end{align}
\end{subequations}
where $P$ denotes the period of \text{CGWSS}, and $\mathbf{\Sigma}_{t}$ and $\mathbf{p}_{t}$ are the periodically varying covariance and PSD, respectively.
\end{definition}

\noindent CGWSS can be considered as a generalized version of GWSS in Definition~\ref{def:gwss} for time-varying graph signals since CGWSS is identical to GWSS when $P=1$.

In this paper, we assume two subgraphs are disconnected\footnote{Technically, this can be done by performing graph clustering for a connected graph.} but the signals on them, $\mathbf{x}_{\text{src},t-1}$ and $\mathbf{x}_{\text{trg},t}$ (please see Fig. \ref{fig:overview}), have similar PSDs:
The similarity is given by the (continuous) kernel function $\tilde{p}_{\text{dom},t}(\lambda)$.
It corresponds to the smoothly interpolated version of the graph filter kernel $p_{\text{dom},t}(\lambda_i):= [\mathbf{p}_{\text{dom},t}]_i$.
Specifically, we assume $\sup_{\lambda}(\tilde{p}_{\text{src},t-1}(\lambda)-\tilde{p}_{\text{trg},t}(\lambda))^{2}<C$ where $C$ is a small constant.
This assumption is required for the scenario where the proposed filter performs alternately between two subgraphs.
We validate that the assumptions practically work well in Section \ref{sec:exp}.

In the following, we formulate a state-space model based on Definition~\ref{def:cgwss}.

\subsection{Control System}

We consider the following state-space model similar to \eqref{eq:kalman_model-state1} and \eqref{eq:kalman_model-observe1}.
\begin{subequations}
\begin{align}
    \mathbf{x}_{\text{dom},t}\!=&T(\mathbf{x}_{\text{dom},t-2})+\mathbf{B}\mathbf{u}_{\text{dom},t-2}+\mathbf{v}_{\text{dom}},
    \label{eq:kalman_model-state_prop}\\
    \mathbf{y}_{\text{dom},t}\!=&\mathbf{C}\mathbf{x}_{\text{dom},t}+\mathbf{w}_{\text{dom}}.\label{eq:kalman_model-observe_prop}
\end{align}
\end{subequations}
where $\mathbf{v}_{\text{dom}}\!\sim\!\mathcal{N}(0,\sigma_{v}^2\mathbf{I})$ and $\mathbf{w}_{\text{dom}}\!\sim\!\mathcal{N}(0,\sigma_{w}^2\mathbf{I})$.
Note that, as illustrated in Fig.~\ref{fig:overview}, we estimate the signal values of the target subgraph at $t$ from those in $t-2$ while we can utilize some (estimated) statistics of the source subgraph at $t-1$.
In \eqref{eq:kalman_model-state_prop}, 
we utilize the transition function $T$, which is not necessarily to be time-invariant, instead of the time-invariant matrix $\mathbf{A}$ in \eqref{eq:kalman_model-state1}. In \eqref{eq:kalman_model-observe_prop}, we utilize the same model as \eqref{eq:kalman_model-observe1}.

\begin{table}[t]
    \caption{Observed and unknown signals. $\text{\cmark}$ and the \textit{blank} mean known and unknown data, respectively. --- represents absent data.}\label{tab:access_signal}
    \centering
    \renewcommand{\arraystretch}{0.85}
    \resizebox{0.5\textwidth}{!}{%
    \begin{tabular}{c|ccc|ccc|ccc}
    \hline
     & $\mathbf{x}_{t-2}$ & $\mathbf{x}_{t-1}$ & $\mathbf{x}_{t}$ & $\mathbf{y}_{t-2}$ & $\mathbf{y}_{t-1}$ & $\mathbf{y}_t$ & $\mathbf{p}_{t-2}$ & $\mathbf{p}_{ t-1}$ & $\mathbf{p}_{t}$ \\ \hline\hline
    src & --- &\cmark & --- & --- & \cmark & --- & --- & \cmark & --- \\ \hline
    trg & \cmark & --- & \  & \cmark & --- & \cmark & \cmark & --- \\ \hline
    \end{tabular}
    }
\end{table}
Table~\ref{tab:access_signal} summarizes which variables are known or unknown.
Our aim is to estimate the unknown signals $\mathbf{x}_{\text{trg},t}$ from known signals.
Note that $\mathbf{\Sigma}_{\text{trg},t}$, which corresponds to $\mathbf{\Sigma}_2$ in \eqref{eq:special_ot} and $T(\cdot)$ in \eqref{eq:kalman_model-state_prop}, is unknown.
Therefore, we first estimate $\mathbf{p}_{\text{trg},t}$ from $\mathbf{p}_{\text{src},t-1}$.

In the standard Kalman filter introduced in Section~\ref{sec:pre}-\ref{sec:kalman-filter}, its estimator is derived in one cluster.
In contrast, we perform the estimation alternately in time between two subgraphs. 
In the following, we derive the proposed Kalman filter.

\subsection{Kalman Filter Transfer}
We derive the cooperative Kalman filter between two subgraphs based on \eqref{eq:kalman_model-state_prop} and \eqref{eq:kalman_model-observe_prop}. For brevity, we replace the subscripts $\cdot_{\text{trg},t-2}$ and $\cdot_{\text{trg},t}$, with $\cdot_1$ and $\cdot_2$, respectively.

Following from Definition~\ref{def:cgwss}, $\mathbf{\Sigma}_{1}$ and $\mathbf{\Sigma}_{2}$ can be jointly diagonalized. Therefore, the RHS in \eqref{eq:special_ot} in the target domain can be rewritten as follows:
\begin{equation}\label{eq:special_ot_2}
 \begin{split}
    T(\mathbf{x}_{1})
    =&\bm{\mu}_{2}
    +\mathbf{Q}(\mathbf{x}_{1}-\bm{\mu}_{1}),
 \end{split}
\end{equation}
where $\mathbf{Q}=\mathbf{U}_{\text{trg}}\text{diag}(\mathbf{p}_{2})\text{diag}(\mathbf{p}_{1})^{-1}\mathbf{U}_{\text{trg}}^\top$.

\begin{figure*}
    \centering
    \includegraphics[width=0.83 \linewidth]{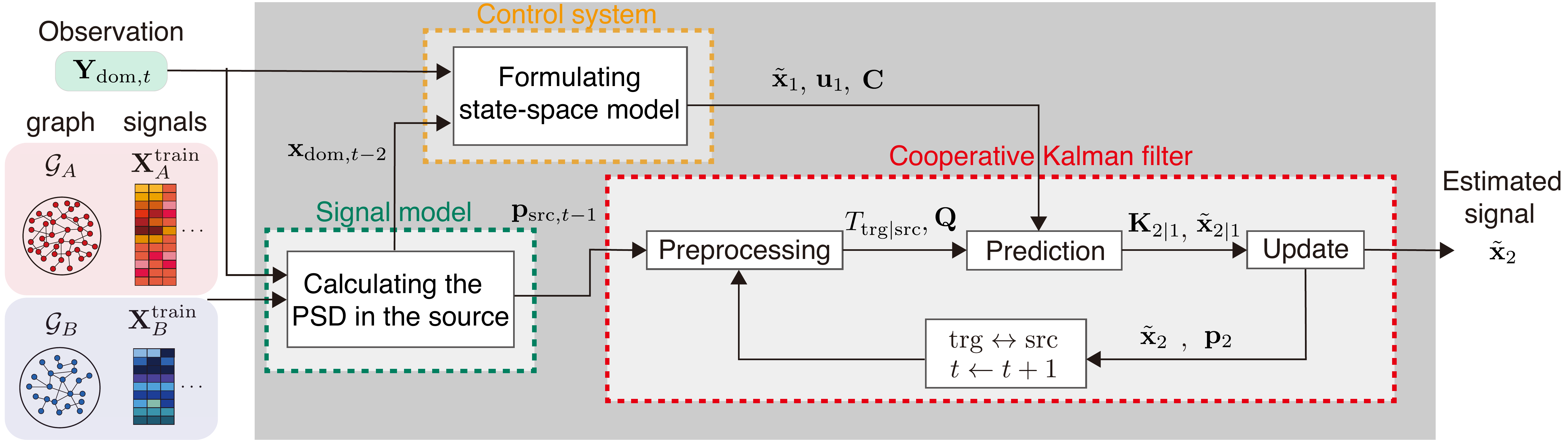}
    \caption{Overview of the time-varying graph signal estimation.}
    \label{fig:outline}
\end{figure*}

To calculate the RHS in \eqref{eq:special_ot_2}, the current PSD $\mathbf{p}_2$ ($=\mathbf{p}_{\text{trg},t}$) is required but unknown (see Table \ref{tab:access_signal}).
Therefore, we estimate $\mathbf{p}_2$ from $\mathbf{p}_{\text{src},t-1}$ by using the graph filter transfer method introduced in Section~\ref{sec:related}.
Consequently, the control laws of the proposed Kalman filter are described as follows and summarized in Algorithm~\ref{alg:CPKF}.
We also illustrate them in Fig.~\ref{fig:outline}.

\begin{description}\label{eq:control-laws}
    \item[Preprocessing step]\text{}
        \begin{enumerate}
            \item Estimating $\mathbf{p}_2$ from $\mathbf{p}_{\text{src},t-1}$ by using the graph filter transfer method in Sec.~\ref{sec:related}.
            \item Calculating the optimal transport map in \eqref{eq:special_ot_2} by using the estimated $\mathbf{p}_2$. We denote the transport map as $T_{\text{trg}|\text{src}}(\tilde{\mathbf{x}}_1)$.
        \end{enumerate}
    
    \item[Prediction step]\text{}
        \begin{enumerate}
            \item Calculating the prior estimation of signals.
            \begin{equation}
                \tilde{\mathbf{x}}_{2|1}=T_{\text{trg}|\text{src}}(\tilde{\mathbf{x}}_{1})+\mathbf{B}\mathbf{u}_{1}.\label{eq:pre_state_prop}
            \end{equation}
            \item Determining the prior error covariance matrix.
            \begin{equation}
                \mathbf{P}_{2|1}=
                \mathbf{Q}\mathbf{P}_{1}\mathbf{Q}^\top
                +\sigma_{v}^2\mathbf{I}.\label{eq:pre_cov_prop}
            \end{equation}
            \item Deriving the Kalman gain.
            \begin{equation}
                \mathbf{K}_{2|1}=\mathbf{P}_{2|1}\mathbf{C}^\top(\mathbf{C}\mathbf{P}_{2|1}\mathbf{C}^\top+\sigma_{w}^2\mathbf{I})^{-1}.\label{eq:kalman_gain_prop}
            \end{equation}
        \end{enumerate}
    \item[Update step]\text{}
        \begin{enumerate}
            \item Estimating the current signals using the Kalman gain.
            \begin{equation}
                \tilde{\mathbf{x}}_{2}=\tilde{\mathbf{x}}_{2|1}+\mathbf{K}_{2|1}(\mathbf{y}_{2}-\mathbf{C}\tilde{\mathbf{x}}_{2|1}).\label{eq:kalman_est_prop}
            \end{equation}
            \item Updating the posterior error covariance matrix.
            \begin{equation}
                \mathbf{P}_{2}=(\mathbf{I}-\mathbf{K}_{2|1}\mathbf{C})\mathbf{P}_{2|1}.\label{eq:post_cov_prop}
            \end{equation}
            \item Swapping \text{src} with \text{trg} and returning to the preprocessing step with $t\leftarrow t+1$.
        \end{enumerate}
\end{description}
\noindent In the initial estimation for each subgraph, $\mathbf{P}_{1}$ in \eqref{eq:pre_cov_prop} is set to $\mathbf{P}_{1}=\delta\mathbf{I}$, similar to the setting in Sec.~\ref{sec:pre}-\ref{sec:kalman-filter}.

\begin{algorithm}
\DontPrintSemicolon
\setlength{\abovedisplayskip}{0pt}
\setlength{\belowdisplayskip}{0pt}
\KwInput{$\tilde{\mathbf{x}}_1$, $\mathbf{u}_1$, $\mathbf{p}_{\text{src},t-1}$, $\mathbf{p}_1$, $\mathbf{P}_1$, $\mathbf{B}$, $\mathbf{C}$, $\sigma_{v}$, $\sigma_{w}$}
{\textbf{Preprocessing}\\
\Indpp 
Estimate $\mathbf{p}_2$ from $\mathbf{p}_{\text{src},t-1}$ by the graph filter transfer in Sec.~\ref{sec:related}\\
Calculate $T_{\text{trg}|\text{src}}(\tilde{\mathbf{x}}_1)$ and $\mathbf{Q}$ in \eqref{eq:special_ot_2}\\
\Indmm\textbf{Prediction}\\
\Indpp$\tilde{\mathbf{x}}_{2|1}=T_{\text{trg}|\text{src}}(\tilde{\mathbf{x}}_{1})+\mathbf{B}\mathbf{u}_{1}$\\
$\mathbf{P}_{2|1}=\mathbf{Q}\mathbf{P}_{1}\mathbf{Q}^\top+\sigma_{v}^2\mathbf{I}$\\
$\mathbf{K}_{2|1}=\mathbf{P}_{2|1}\mathbf{C}^\top(\mathbf{C}\mathbf{P}_{2|1}\mathbf{C}^\top+\sigma_{w}^2\mathbf{I})^{-1}$\\
\Indmm\textbf{update}\\
\Indpp$\tilde{\mathbf{x}}_{2}=\tilde{\mathbf{x}}_{2|1}+\mathbf{K}_{2|1}(\mathbf{y}_{2}-\mathbf{C}\tilde{\mathbf{x}}_{2|1})$\\
$\mathbf{P}_{2}=(\mathbf{I}-\mathbf{K}_{2|1}\mathbf{C})\mathbf{P}_{2|1}$\\
\Indmm $t\leftarrow t+1$ with swapping $\text{src}$ with $\text{trg}$
}\\
\KwOutput{$\tilde{\mathbf{x}}_2$, $\mathbf{p}_2$, $\mathbf{P}_2$}
\caption{Control laws of cooperative KF}\label{alg:CPKF}
\end{algorithm}

\section{EXPERIMENTS}\label{sec:exp} 
In this section, we perform signal estimation experiments for synthetic and real-world data.

\subsection{Baseline methods and performance measure}
Since there is no prior work on the cooperative Kalman filter to the best of our knowledge, we use the following well-known methods as baseline methods. 
\begin{description}
    \item[Ridge regression with Tikhonov regularization (RRTK): ]The first baseline is signal estimation based on Tikhonov regularization \cite{kobak2020optimal}.
    The estimated signal can be written as: 
    \begin{equation}
    \begin{split}\label{Tikhonov1}
        \tilde{\mathbf{x}}^{\text{RRTK}}_{2}=&\argmin_{\mathbf{x}_{2}}\|\mathbf{y}_{2}-\mathbf{C}\mathbf{x}_{2}\|^2_2
        +\zeta\mathbf{x}_{2}^\top\mathbf{L}_{\text{trg}}\mathbf{x}_{2}\\
        =&(\mathbf{C}^\top \mathbf{C}+\zeta\mathbf{L}_{\text{trg}})^{-1}\mathbf{C}^\top\mathbf{y}_{2},
        \end{split}
    \end{equation}
    where $\zeta$ is a parameter, which controls the intensity of the regularization term and set to $\zeta=0.05$.
    This method only uses the space equation, and the signals are neither GWSS nor CGWSS. In contrast, the proposed method uses the state-space model, and the signals are CGWSS.
    \item[Transferred graph Wiener filter (TGWF): ]
    The second one is signal estimation based on a transferred graph Wiener filter, as mentioned in Section~\ref{sec:related} \cite{yamada_graph_2022}. 
    The estimated signal can be written as \cite{yamada_graph_2022}:
    \begin{equation}
        \tilde{\mathbf{x}}^{\text{TGWF}}_{2}=\mathbf{H}\mathbf{y}_2+\mathbf{b},
    \end{equation}
    where $\mathbf{H}=\mathbf{\Sigma}_2\mathbf{C}^\top(\mathbf{C}\mathbf{\Sigma}_2\mathbf{C}^\top)^{-1}, \quad \mathbf{b}=(\mathbf{I}-\mathbf{H}\mathbf{C})\bm{\mu}_2$.
    The covariance matrix $\mathbf{\Sigma}_2$ is calculated from the estimated PSD $\mathbf{p}_{2}$ (see Definition~\ref{def:gwss} and Section~\ref{sec:related}).
\end{description}

In the following experiments, we set $\sigma_{v}=0$ in \eqref{eq:kalman_model-state_prop} to conduct the experiments for a fair comparison.
We consider additive white Gaussian noise on observations with three different standard deviations $\sigma_{w}$.

We evaluate the estimation performance of each method with the MSE for each $\sigma_{w}$.

\subsection{Synthetic Graph Signals}
We compare the estimation accuracy of the proposed Kalman filter on synthetic data with that of alternative methods.

\subsubsection{Dataset}
We construct two different random sensor (RS) graphs\footnote{Random sensor graphs are implemented by $k$ nearest neighbor graphs whose nodes are randomly distributed in 2-D space $[0,1]\times[0,1]$ (See \cite{perraudin2014gspbox}).}, $\mathcal{G}_{A}$ and $\mathcal{G}_{B}$ with $N_A=90$ and $N_B=45$, as subgraphs. 
The period of CGWSS signals is set to $P=8$ and their PSDs $\{\mathbf{p}_{{\text{dom}},p}\}_{p=0,\ldots,P-1}$ are given by four different low-pass filters as functions in $\mathbf{L}_{\text{dom}}$:
\begin{equation}
[\mathbf{p}_{{\text{dom}},p}]_i=
    \begin{cases}
         1-\lambda_{i}/\lambda_{\text{max}} &(p=0,4), \\
        \exp(-\lambda_{i}/\lambda_{\text{max}}) &(p=1,5), \\
        1/(1+\lambda_{i}) &(p=2,6),\\
        \cos(\pi\lambda_{i}/2\lambda_{\max}) &(p=3,7).
    \end{cases}
\end{equation}
Accordingly, we generate samples of the signals conforming to $\mathcal{N}(\mathbf{1},\mathbf{U}_{\text{dom}} \text{diag}(\mathbf{p}_{\text{dom},p})\mathbf{U}_{\text{dom}}^\top)$, where $p=t (\text{mod}\,P)$ for $t=1,\ldots,T$.
We denote the training and test datasets by $\mathbf{X}_{\text{dom}}^{\text{train}}\in\mathbb{R}^{N_{\text{dom}} \times T_{\text{train}}}$ and $\mathbf{X}_{\text{dom}}^{\text{test}}\in\mathbb{R}^{N_{\text{dom}} \times T_{\text{test}}}$, consisting of $T_{\text{train}} = 200$ and $T_{\text{test}} = 40$ samples, respectively. 

We consider a data update process for the sequential signal estimation experiment.
Let $\mathbf{X}^{(l)}_{\text{dom},p}\in\mathbb{R}^{N_{\text{dom}}\times K}$ be a data slot at $p$, where $K=T_{\text{train}}/P$ and $l$ indicates the $l$th cycle, i.e., it satisfies $t=lP+p$.
We update the data slot at every time instance in a warm-start manner:
\begin{equation}
    \mathbf{X}_{\text{src},p}^{(l+1)}=
    \begin{cases}
        \mathbf{X}_{\text{src},p}^{\text{train}} & \text{if }l=0,\\
        \left[\tilde{\mathbf{x}}_2,\left[\mathbf{X}_{\text{src},p}^{(l)}\right]_{:,1:K-1}\right] & \text{otherwise},
    \end{cases}\label{eq:data_slot}
\end{equation}
where 
$\left[\mathbf{X}_{\text{src},p}^{(l)}\right]_{:,1:K-1}$ denotes the submatrix of $\mathbf{X}_{\text{src},p}^{(l)}$ whose columns from $1$ to $K-1$.
In \eqref{eq:data_slot}, We divide $\mathbf{X}_{\text{dom}}^{\text{train}}$ into $P$ periods and set them as the initial values for each period. 
 \subsubsection{Experimental Setup}

At the initial estimation for each subgraph, i.e., $t = 1,2$, we set the scaling factor of the error covariance, $\mathbf{P}_1$ in \eqref{eq:pre_cov_prop} to $\delta = 1$, and the estimation of signal $\tilde{\mathbf{x}}_1 = \left[{\mathbf{X}}_{\text{trg},p}^{\text{train}} \right]_{:1}$ in \eqref{eq:pre_state_prop}.
In \eqref{eq:pre_state_prop}, we set the input matrix $\mathbf{B}=\mathbf{I}$.
We use a proportional (PI) control strategy for the control input \cite{10.5555/516039}: 
    $\mathbf{u}_{1}=\eta(\tilde{\mathbf{x}}_{1} - \bar{\mathbf{x}}_{\text{trg},p})$,
where $\eta$ is a factor of proportionality and $\bar{\mathbf{x}}_{\text{trg},p}$ is a mean vector in the latest data slot.
We empirically set $\eta=5.0\times 10^{-2}$.
We use a downsampling operator $\mathbf{C}=\mathbf{I}_{\mathcal{M}}$ as the observation matrix in \eqref{eq:kalman_model-observe_prop}, where $\mathbf{I}_{\mathcal{M}}$ is the submatrix of $\mathbf{I}$ whose rows are extracted according to the subset of nodes $\mathcal{M}$. 
We randomly sample the signals with $M_{A}=85$ and $M_{B}=43$ in \eqref{eq:kalman_model-observe_prop} for $\mathcal{G}_A$ and $\mathcal{G}_B$, respectively.
\begin{figure*}
    \centering
    \includegraphics[width=0.9 \linewidth]{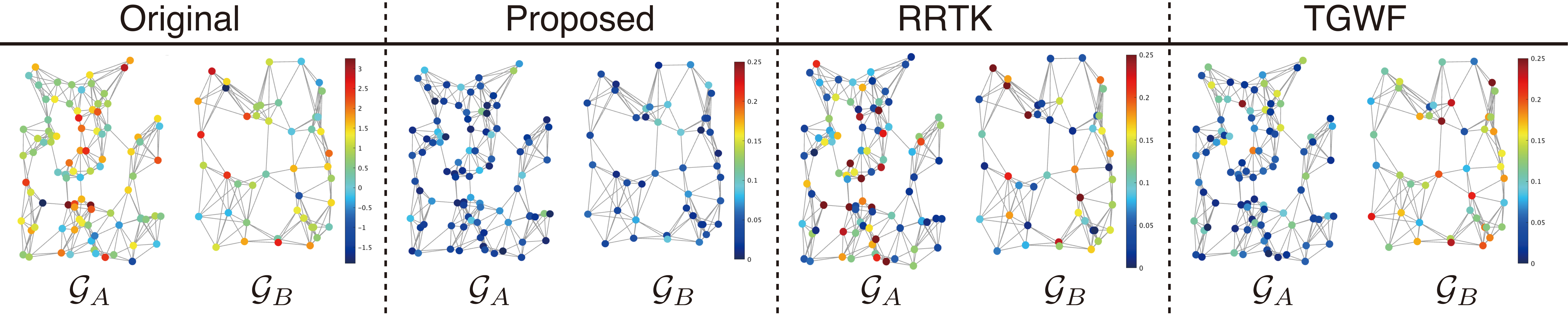}
    \caption{Signal estimation experiments for noisy graph signals on two different RS graphs with $N_A= 90 $, $N_B = 45$ nodes at time $t=8$, $t=7$, respectively. The color of a node represents the absolute error between original and estimated signals.}
    \label{fig:synthe_result}
\end{figure*}

\subsubsection{Results}
\begin{figure}[t]
 \centering
   \subfloat{\includegraphics[width=0.7\linewidth]{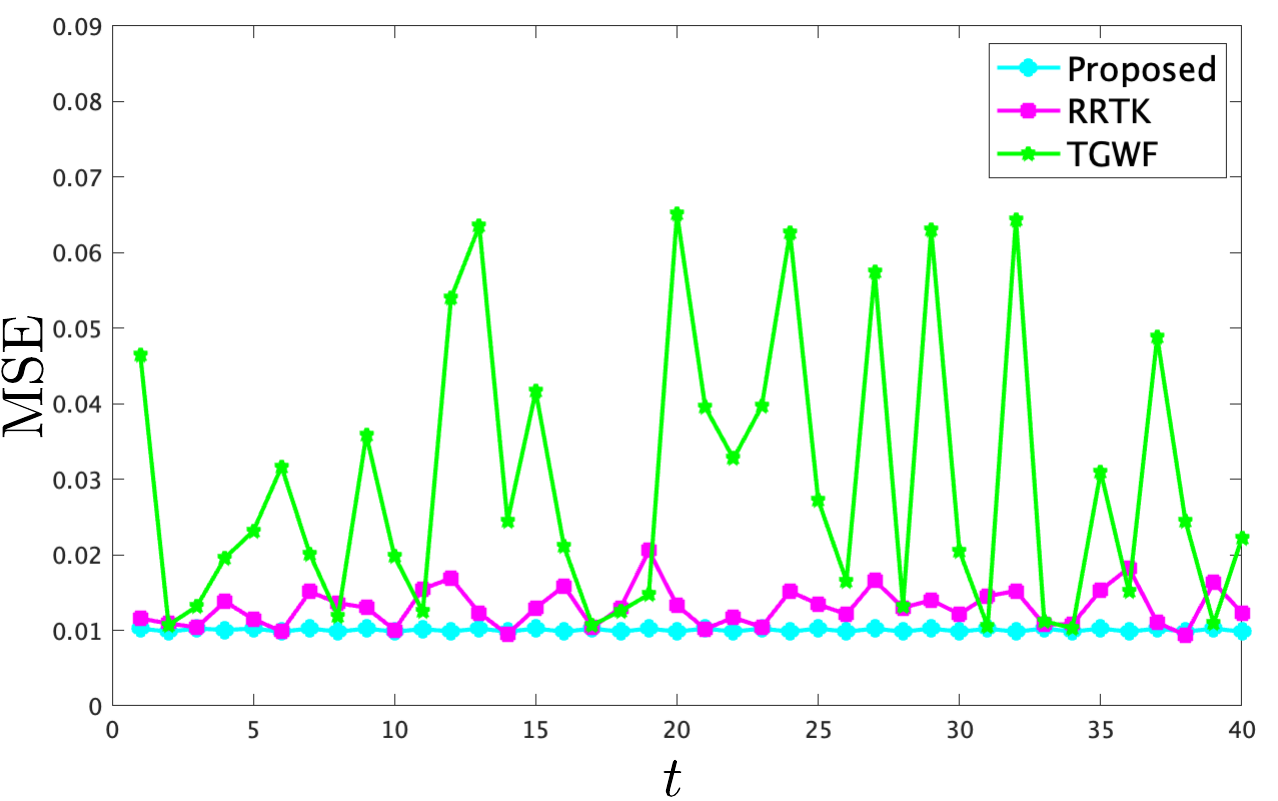}}
     \caption{Comparison of signal estimation MSEs for the synthetic data ($\sigma_{w}=0.10$). The MSEs for each method is plotted as a holizeontal line.}
     \label{fig:mse_time_synth}
      
 \end{figure}

Table~\ref{tab:synth} summarizes the average MSEs. 
We also visualize an example of absolute errors between the original and estimated signals in Fig.~\ref{fig:synthe_result}.
Fig.~\ref{fig:mse_time_synth} plots MSEs over time.

In Table \ref{tab:synth}, the proposed method outperforms alternative methods for all $\sigma_{w}$. In Fig.~\ref{fig:mse_time_synth}, we observe that the proposed method shows consistent estimation performance for all $t$, while those of the other methods oscillate significantly over time. This is because the proposed method can compensate its own estimation using signals on the previous time instance in the same subgraph and that on the different subgraph with similar statistics,
while other methods perform the estimation independently at every time instance.
\begin{table}[t]
    \centering
    \caption{Experimental results on synthetic dataset.}
    \renewcommand{\arraystretch}{0.8}
    \begin{tabular}{c|c|c|c}
    \hline
          \multirow{2}{*}{$\sigma_{w}$} &  \multicolumn{3}{c}{Average MSE($10^{-2}$)} \\\cline{2-4} & Proposed &  RRTK  & TGWF \\ \hline\hline
         $0.05$ & $\mathbf{0.33}$  &$0.92$ & $2.38$  \\ \hline
         $0.10$ & $\mathbf{1.00}$ & $1.31$ & $2.94$  \\ \hline
         $0.15$ & $\mathbf{2.26}$ & $2.34$ & $4.29$  \\ \hline
    \end{tabular}
    \label{tab:synth}
\end{table}

\subsection{Real-World Data}
Here, we validate the proposed method with several real-world datasets.

\begin{figure*}
    \centering
    \includegraphics[width = 0.89 \linewidth]{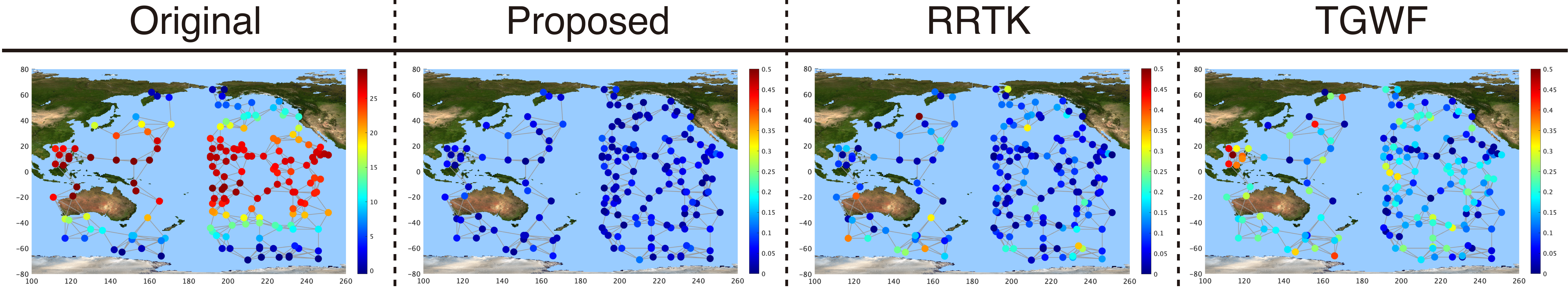}
    \caption{Signal estimation experiments on sea surface temperature data. The color of a node indicates the absolute error between the original and estimated signals. The right-side cluster and left-side cluster correspond to $\mathcal{G}_A$ and $\mathcal{G}_B$ at $t = 24,$ and $t= 23$, respectively.}
    \label{fig:real_result}
\end{figure*}

\begin{figure*}[t]
 \centering
   \subfloat{\includegraphics[width=0.29\linewidth]{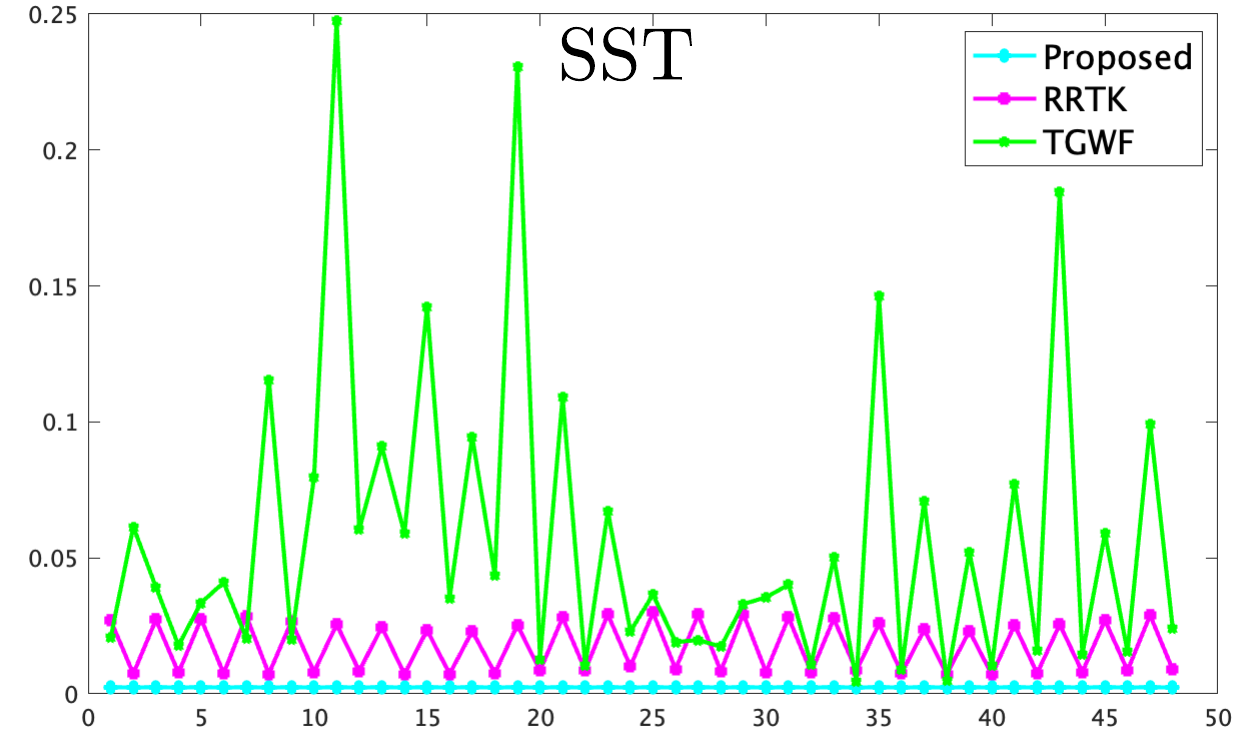}}
   \subfloat{\includegraphics[width=0.29\linewidth]{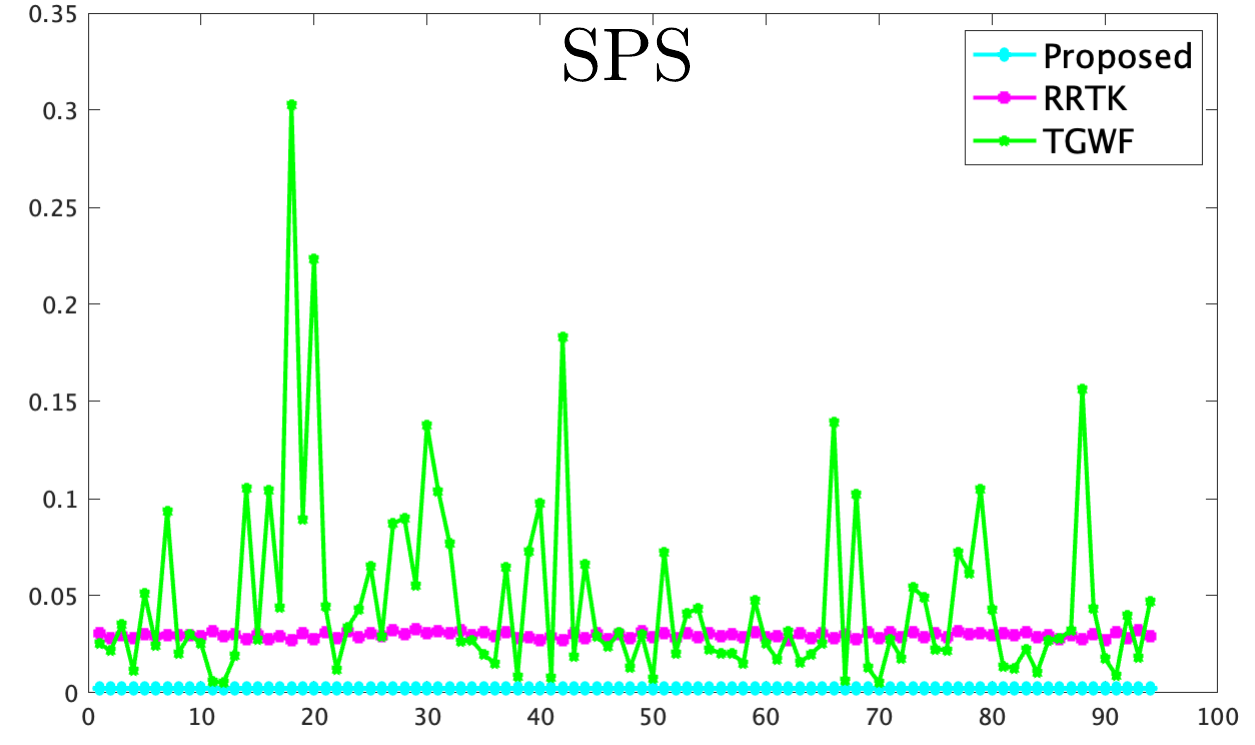}}
   \subfloat{\includegraphics[width=0.29\linewidth]{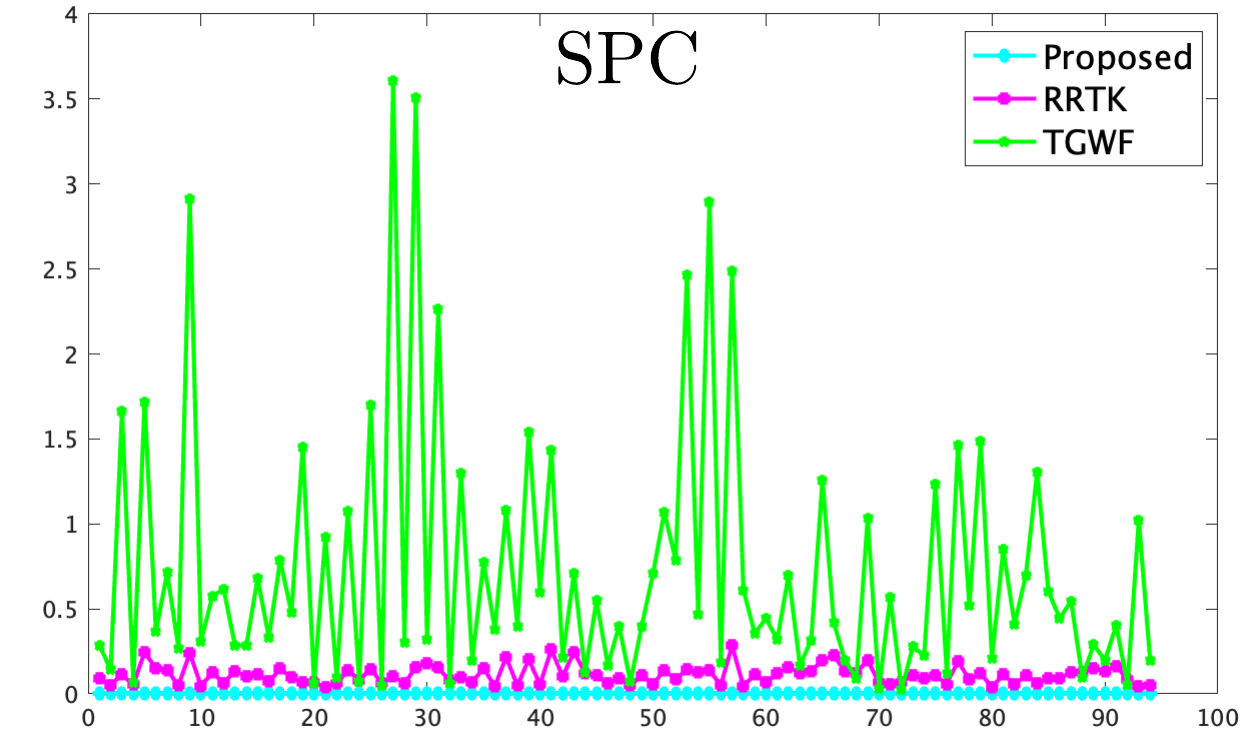}}
     \caption{Comparison of signal estimation MSEs for real-world data ($\sigma_{w}=0.05$). The average MSE for each method is plotted as a holizeontal line.}
     \label{fig:mse_time_inst}
 \end{figure*}

\subsubsection{Datasets}
We use three different environmental sensor data summarized as follows.
\begin{description}
    \item[Sea surface temperature data (SST) \cite{rayner2003global}: ]It records snapshots of global SSTs (\si{\degreeCelsius}) for every month from 2004 to 2021. Each SST is observed at intersections of 1-degree latitude-longitude grids.
    \item[Surface pressure data (SPS)\cite{greene2019climate}: ]It records snapshots of global surface pressure (\si{\kilo\pascal/\metre^{2}})\footnote{Surface pressure is the pressure observed at ground level.} for every month from 1978 to 2024.
    Every SPS is sampled at intersections of $94$-lines of latitude and of $193$-lines longitude.
    \item[Surface precipitation data (SPC)\cite{greene2019climate}: ]It records snapshots of global SPCs (\si{\milli\metre/\metre^{2}}) for every month from 1978 to 2024. Every SPC is sampled at intersections of $94$-lines of latitude and $193$-lines of longitude.
\end{description}

\subsubsection{Experimental Setup}

We extract the datasets from the above three real-world data over consecutive $T$ months. We use snapshots from the first $T_{\text{train}}$ months within $T$ months for training, and the remaining $T_{\text{test}}$ months for test.
We randomly sample two disjoint sets of nodes, $N_A$, $N_B$, from the measured intersections within two distinct latitude ranges. Then, we construct two RS graphs, $\mathcal{G}_A$, $\mathcal{G}_B$, for each set as subgraphs. We assume the period $P$ is consistent across the two graphs and estimate $P$ by using a period estimation method \cite{mcneil_temporal_2021,tenneti_nested_2015}. 
We estimate the initial PSD from the training data by using a PSD estimation method \cite{perraudin_stationary_2017,perraudin2014gspbox}. The observation matrix in \eqref{eq:kalman_model-observe_prop} is set to a downsampling operator $\mathbf{I}_{\mathcal{M}}$. We randomly sample the signals with a node sampling ratio $M / N=0.95$. 
We set the scaling factor of the initial error covariance $\delta = 1$, and the initial estimation of signal $\tilde{\mathbf{x}}_1 = \left[{\mathbf{X}}_{\text{trg},p}^{\text{train}} \right]_{:1}$ in \eqref{eq:pre_state_prop}.
For the control input, we use PI control strategy with the same setting as the preceding experiment. 
We summarize the parameters of the experimental setup for each dataset in Table~\ref{tab:param}.

\begin{table}[t]
    \caption{Parameters for each real-world datasets}\label{result1}
    \centering
    \renewcommand{\arraystretch}{0.75}
    \begin{tabular}{l|c|c|c|c|c|c|c|c}
    \hline
        & $N_{A}$    &  $N_{B}$ & $P$ & $K$ & $T$ 
        &$T_{\text{train}}$ & $T_{\text{test}}$& $\zeta$ \\ \hline \hline
    \textbf{SST}\cite{rayner2003global} & 123 &  46  & 12 & 14 & 216   & 168 & 48 & 0.01\\\hline 
    \textbf{SPS}\cite{greene2019climate} & 90 &  45 & 47 & 8 & 470 & 376  & 94 &0.01 \\\hline 
    \textbf{SPC}\cite{greene2019climate} & 90 &  45 & 47 & 8 & 470 & 376 & 94 & 0.05\\\hline 
    \end{tabular}
    \label{tab:param}
\end{table}

\subsubsection{Results}
Table~\ref{tab:result} summarizes the average MSE for each real-world data and Fig.~\ref{fig:mse_time_inst} plots MSEs over time. Fig.~\ref{fig:real_result} visualizes an example of absolute errors between the original and estimated signals in the experiments with SST data. 
\begin{table}[t]
\caption{Experimental results on real-world datasets.}
    \centering
    \renewcommand{\arraystretch}{0.8}
    \begin{tabular}{c|c|c|c}
        \hline
         \multirow{2}{*}{$\sigma_{w}$} &  \multicolumn{3}{c}{Average MSE ($\times10^{-1}$)} \\\cline{2-4} & Proposed & RRTK & TGWF \\ \hline\hline
        \multicolumn{4}{c}{Sea surface temperature (SST) }\\ \hline\hline
         0.05 &  $\bm{0.03}$ & 0.19 & 0.50 \\\hline
         0.10 &  $\bm{0.11}$ & 0.28 & 0.57 \\ \hline
         0.15 &  $\bm{0.24}$ & 0.41 & 0.69 \\\hline \hline        \multicolumn{4}{c}{Surface pressure (SPS)}\\ \hline\hline
         0.05 &  $\bm{0.03}$ & 0.30 & 0.44  \\\hline
         0.10 &  $\bm{0.10}$ & 0.38 & 0.56 \\ \hline
         0.15 &  $\bm{0.24}$ & 0.52 & 0.70 \\\hline\hline
        \multicolumn{4}{c}{Surface precipitation (SPC)}\\ \hline\hline
         0.05 &  $\bm{0.02}$ & 1.20 & 7.34 \\\hline
         0.10 &  $\bm{0.10}$ & 1.11 & 8.32 \\ \hline
         0.15 &  $\bm{0.22}$ & 1.21 & 8.70 \\\hline

            \end{tabular}
    \label{tab:result}
\end{table}

As seen in Table \ref{tab:result}, the proposed method outperforms alternative methods for all $\sigma_{w}$. 
In Fig. \ref{fig:mse_time_inst}, it is observed that the estimation errors of two alternative methods oscillate significantly over time, while those of the proposed method are consistently low. Particularly, in Fig.~\ref{fig:real_result}, the estimated SSTs of the alternative methods show large errors in the latitude ranges from $-80$\si{\degree} to $-40$\si{\degree} and from $40$\si{\degree} to $80$\si{\degree}.
Note that, in these areas, several sea currents having large temperature differences (e.g., Kuroshio current and Oyashio current) simultaneously exist.
Our method can successfully compensate such a complex behavior by using signals on the previous time instance in the same subgraph and the statistics on the different subgraph in the current time instance.


\section{CONCLUSION}\label{sec:conclude}
In this paper, we propose a Kalman filter transfer method for estimating time-varying signals among multiple subgraphs.
Initially, we assume that the time-varying graph signals conform to CGWSS for both subgraphs. We then formulate the control system based on the state-space model. In the state equation, we describe the temporal evolution of CGWSS signals using optimal transport. To alternately perform signal estimation over time in the source and target, we transfer the set of parameters in the Kalman filter from the source to the target. Our experiments demonstrate that the proposed method effectively estimates the time-varying graph signals for clustered networks.

\bibliography{reference}

\begin{thebibliography}{10}

\bibitem{chong2003sensor}
C.-Y. Chong and S.~P. Kumar, ``Sensor networks: evolution, opportunities, and challenges,'' {\em Proc. IEEE}, vol.~91, no.~8, pp.~1247--1256, 2003.

\bibitem{lee2011discovering}
W.-H. Lee, S.-S. Tseng, J.-L. Shieh, and H.-H. Chen, ``Discovering traffic bottlenecks in an urban network by spatiotemporal data mining on location-based services,'' {\em IEEE Trans. Intell. Transp. Syst.}, vol.~12, no.~4, pp.~1047--1056, 2011.

\bibitem{martini2015automatic}
A.~Martini, M.~Troncossi, and A.~Rivola, ``Automatic leak detection in buried plastic pipes of water supply networks by means of vibration measurements,'' {\em Shock Vib.}, vol.~2015, no.~1, p.~165304, 2015.

\bibitem{gabrys2000general}
B.~Gabrys and A.~Bargiela, ``General fuzzy min-max neural network for clustering and classification,'' {\em IEEE Trans. Neural Netw.}, vol.~11, no.~3, pp.~769--783, 2000.

\bibitem{van_der_merwe_square-root_2001}
R.~Van Der~Merwe and E.~Wan, ``The square-root unscented kalman filter for state and parameter-estimation,'' in {\em Proc. {IEEE} Int. Conf. Acoust. Speech Signal Process. (ICASSP)}, vol.~6, pp.~3461--3464, {IEEE}, 2001.

\bibitem{clarke1989properties}
D.~W. Clarke and C.~Mohtadi, ``Properties of generalized predictive control,'' {\em Automatica}, vol.~25, no.~6, pp.~859--875, 1989.

\bibitem{kouvaritakis2016model}
B.~Kouvaritakis and M.~Cannon, ``Model predictive control,'' {\em Classical, Robust and Stochastic.}, vol.~38, pp.~13--56, 2016.

\bibitem{maciejowski2007predictive}
J.~M. Maciejowski and M.~Huzmezan, ``Predictive control,'' in {\em Robust Flight Control: A Design Challenge}, pp.~125--134, Springer, 2007.

\bibitem{sutton1981toward}
R.~S. Sutton and A.~G. Barto, ``Toward a modern theory of adaptive networks: expectation and prediction.,'' {\em Psychological review}, vol.~88, no.~2, p.~135, 1981.

\bibitem{sagi2023extended}
G.~Sagi, N.~Shlezinger, and T.~Routtenberg, ``Extended kalman filter for graph signals in nonlinear dynamic systems,'' in {\em Proc. {IEEE} Int. Conf. Acoust. Speech Signal Process. (ICASSP)}, pp.~1--5, IEEE, 2023.

\bibitem{ramezani2018joint}
M.~Ramezani-Mayiami and B.~Beferull-Lozano, ``Joint graph learning and signal recovery via kalman filter for multivariate auto-regressive processes,'' in {\em Proc. Eur. Signal Process. Conf. (EUSIPCO)}, pp.~907--911, IEEE, 2018.

\bibitem{bishop2001introduction}
G.~Bishop, G.~Welch, {\em et~al.}, ``An introduction to the kalman filter,'' {\em Proc of SIGGRAPH, Course}, vol.~8, no.~27599-23175, p.~41, 2001.

\bibitem{li1989state}
S.~Li, K.~Y. Lim, and D.~G. Fisher, ``A state space formulation for model predictive control,'' {\em AIChE J.}, vol.~35, no.~2, pp.~241--249, 1989.

\bibitem{zhu2015green}
C.~Zhu, V.~C. Leung, L.~Shu, and E.~C.-H. Ngai, ``Green internet of things for smart world,'' {\em IEEE Access}, vol.~3, pp.~2151--2162, 2015.

\bibitem{natali_learning_2022}
A.~Natali, E.~Isufi, M.~Coutino, and G.~Leus, ``Learning time-varying graphs from online data,'' {\em {IEEE} Open J. Signal Process.}, vol.~3, pp.~212--228, 2022.

\bibitem{pan_survey_2010}
S.~J. Pan and Q.~Yang, ``A survey on transfer learning,'' {\em {IEEE} Trans. Knowl. Data Eng.}, vol.~22, no.~10, pp.~1345--1359, 2010.

\bibitem{shuman2013emerging}
D.~I. Shuman, S.~K. Narang, P.~Frossard, A.~Ortega, and P.~Vandergheynst, ``The emerging field of signal processing on graphs: Extending high-dimensional data analysis to networks and other irregular domains,'' {\em IEEE Signal Process. Mag.}, vol.~30, no.~3, pp.~83--98, 2013.

\bibitem{ortega_graph_2018}
A.~Ortega, P.~Frossard, J.~Kova{\v{c}}evi{\'c}, J.~M. Moura, and P.~Vandergheynst, ``Graph signal processing: Overview, challenges, and applications,'' {\em Proc. IEEE}, vol.~106, no.~5, pp.~808--828, 2018.

\bibitem{perraudin_stationary_2017}
N.~Perraudin and P.~Vandergheynst, ``Stationary signal processing on graphs,'' {\em {IEEE} Trans. Signal Process.}, vol.~65, no.~13, pp.~3462--3477, 2017.

\bibitem{hara_graph_2023}
J.~Hara, Y.~Tanaka, and Y.~C. Eldar, ``Graph signal sampling under stochastic priors,'' {\em {IEEE} Trans. Signal Process.}, vol.~71, pp.~1421--1434, 2023.

\bibitem{marques_stationary_2017}
A.~G. Marques, S.~Segarra, G.~Leus, and A.~Ribeiro, ``Stationary graph processes and spectral estimation,'' {\em {IEEE} Trans. Signal Process.}, vol.~65, no.~22, pp.~5911--5926, 2017.

\bibitem{villani2009optimal}
C.~Villani {\em et~al.}, {\em Optimal transport: old and new}, vol.~338.
\newblock Springer, 2009.

\bibitem{delon_wasserstein-type_2020}
J.~Delon and A.~Desolneux, ``A wasserstein-type distance in the space of gaussian mixture models,'' {\em SIAM J. Imaging Sci.}, vol.~13, no.~2, pp.~936--970, 2020.

\bibitem{yamada_graph_2022}
K.~Yamada, ``Graph filter transfer via probability density ratio weighting,'' {\em arXiv:2210.14633}, 2022.

\bibitem{anderson2005optimal}
B.~D. Anderson and J.~B. Moore, {\em Optimal filtering}.
\newblock Courier Corporation, 2005.

\bibitem{mesbah2016stochastic}
A.~Mesbah, ``Stochastic model predictive control: An overview and perspectives for future research,'' {\em IEEE Control Syst. Mag.}, vol.~36, no.~6, pp.~30--44, 2016.

\bibitem{santambrogio2015optimal}
F.~Santambrogio, {\em Optimal transport for applied mathematicians}, vol.~55.
\newblock Springer, 2015.

\bibitem{villani2009wasserstein}
C.~Villani {\em et~al.}, ``The wasserstein distances,'' in {\em Optimal Transport: Old and New}, vol.~338, pp.~93--111, Springer, 2009.

\bibitem{panaretos2019statistical}
V.~M. Panaretos and Y.~Zemel, ``Statistical aspects of wasserstein distances,'' {\em Annu. Rev. Stat. Its Appl.}, vol.~6, pp.~405--431, 2019.

\bibitem{bjorck1983schur}
{\AA}.~Bj{\"o}rck and S.~Hammarling, ``A schur method for the square root of a matrix,'' {\em Linear Algebra Appl.}, vol.~52, pp.~127--140, 1983.

\bibitem{isufi_autoregressive_2017}
E.~Isufi, A.~Loukas, A.~Simonetto, and G.~Leus, ``Autoregressive moving average graph filtering,'' {\em {IEEE} Trans. Signal Process.}, vol.~65, no.~2, pp.~274--288, 2017.

\bibitem{kroizer_bayesian_2022}
A.~Kroizer, T.~Routtenberg, and Y.~C. Eldar, ``Bayesian estimation of graph signals,'' {\em {IEEE} Trans. Signal Process.}, vol.~70, pp.~2207--2223, 2022.

\bibitem{kay1993statistical}
S.~M. Kay, ``Statistical signal processing: estimation theory,'' {\em Prentice Hall}, vol.~1, pp.~Chapter--3, 1993.

\bibitem{kobak2020optimal}
D.~Kobak, J.~Lomond, and B.~Sanchez, ``The optimal ridge penalty for real-world high-dimensional data can be zero or negative due to the implicit ridge regularization,'' {\em J. Mach. Learn. Res.}, vol.~21, no.~169, pp.~1--16, 2020.

\bibitem{perraudin2014gspbox}
N.~Perraudin, J.~Paratte, D.~Shuman, L.~Martin, V.~Kalofolias, P.~Vandergheynst, and D.~K. Hammond, ``Gspbox: A toolbox for signal processing on graphs,'' {\em arXiv:1408.5781}, 2014.

\bibitem{10.5555/516039}
K.~Ogata, {\em Modern Control Engineering}.
\newblock Prentice Hall PTR, 4th~ed., 2001.

\bibitem{rayner2003global}
N.~Rayner, D.~E. Parker, E.~Horton, C.~K. Folland, L.~V. Alexander, D.~Rowell, E.~C. Kent, and A.~Kaplan, ``Global analyses of sea surface temperature, sea ice, and night marine air temperature since the late nineteenth century,'' {\em J. Geophys. Res. Atmos.}, vol.~108, no.~D14, 2003.

\bibitem{greene2019climate}
C.~A. Greene, K.~Thirumalai, K.~A. Kearney, J.~M. Delgado, W.~Schwanghart, N.~S. Wolfenbarger, K.~M. Thyng, D.~E. Gwyther, A.~S. Gardner, and D.~D. Blankenship, ``The climate data toolbox for matlab,'' {\em Geochem. Geophys. Geosyst.}, vol.~20, no.~7, pp.~3774--3781, 2019.

\bibitem{mcneil_temporal_2021}
M.~J. {McNeil}, L.~Zhang, and P.~Bogdanov, ``Temporal graph signal decomposition,'' in {\em Proc. 27th ACM SIGKDD Conf. Knowl. Discov. Data Min.}, pp.~1191--1201, {ACM}, 2021.

\bibitem{tenneti_nested_2015}
S.~V. Tenneti and P.~P. Vaidyanathan, ``Nested periodic matrices and dictionaries: New signal representations for period estimation,'' {\em {IEEE} Trans. Signal Process.}, vol.~63, no.~14, pp.~3736--3750, 2015.

\end{thebibliography}
\bibliographystyle{ieeetr}

\end{document}